\documentclass[conference]{IEEEtran}
\IEEEoverridecommandlockouts
\usepackage{cite}
\usepackage{amsmath,amssymb,amsfonts}
\usepackage{algorithmic}
\usepackage{graphicx}
\usepackage{textcomp}
\usepackage{xcolor}
\usepackage{listings}
\usepackage{color}
\usepackage{multicol}
\usepackage{verbatim}

\def\BibTeX{{\rm B\kern-.05em{\sc i\kern-.025em b}\kern-.08em
    T\kern-.1667em\lower.7ex\hbox{E}\kern-.125emX}}

\newcommand{\stimes}{{\times}}

\begin{document}


\title{STOMP: A Tool for Evaluation of Scheduling\\
Policies in Heterogeneous Multi-Processors}

\author{
\IEEEauthorblockN{Augusto Vega}
\IEEEauthorblockA{
\textit{IBM Research}\\
Yorktown Heights, NY \\
ajvega@us.ibm.com}
\and
\IEEEauthorblockN{Aporva Amarnath}
\IEEEauthorblockA{
\textit{University of Michigan}\\
Ann Arbor, MI \\
aporvaa@umich.edu}
\and
\IEEEauthorblockN{John-David Wellman}
\IEEEauthorblockA{
\textit{IBM Research}\\
Yorktown Heights, NY \\
wellman@us.ibm.com}
\and
\IEEEauthorblockN{Hiwot Kassa}
\IEEEauthorblockA{
\textit{University of Michigan}\\
Ann Arbor, MI \\
hiwot@umich.edu}
\and
\IEEEauthorblockN{Subhankar Pal}
\IEEEauthorblockA{
\textit{University of Michigan}\\
Ann Arbor, MI \\
subh@umich.edu}
\and
\IEEEauthorblockN{~~~~~~~Hubertus Franke}
\IEEEauthorblockA{
~~~~~~~\textit{IBM Research}\\
~~~~~~~Yorktown Heights, NY \\
~~~~~~~frankeh@us.ibm.com}
\and
\IEEEauthorblockN{Alper Buyuktosunoglu}
\IEEEauthorblockA{
\textit{IBM Research}\\
Yorktown Heights, NY \\
alperb@us.ibm.com}
\and
\IEEEauthorblockN{Ronald Dreslinski}
\IEEEauthorblockA{
\textit{University of Michigan}\\
Ann Arbor, MI \\
rdreslin@umich.edu}
\and
\IEEEauthorblockN{Pradip Bose}
\IEEEauthorblockA{
\textit{IBM Research}\\
Yorktown Heights, NY \\
pbose@us.ibm.com}
}

\maketitle

\thispagestyle{plain}
\pagestyle{plain}

\begin{abstract}
The proliferation of heterogeneous chip multiprocessors in recent years has reached unprecedented levels. Traditional homogeneous platforms have shown fundamental limitations when it comes to enabling high-performance yet-ultra-low-power computing, in particular in application domains with real-time execution deadlines or criticality constraints.
By combining the right set of general purpose cores and hardware accelerators together, along with proper chip interconnects and memory technology, heterogeneous chip multiprocessors have become an effective high-performance and low-power computing alternative.

One of the challenges of heterogeneous architectures relates to efficient scheduling of application tasks (processes, threads) across the variety of options in the chip. As a result, it is key to provide tools to enable early-stage prototyping and evaluation of new scheduling policies for heterogeneous platforms. In this paper, we present STOMP (Scheduling Techniques Optimization in heterogeneous Multi-Processors), a simulator for fast implementation and evaluation of task scheduling policies in multi-core/multi-processor systems with a convenient interface for ``plugging'' in new scheduling policies in a simple manner. Thorough validation of STOMP exhibits small relative errors when compared against closed-formed equivalent models during steady-state analysis.
\\
\end{abstract}


\section{Introduction}
\label{introduction}

With the imminent ``end'' of Moore's Law, recent years have witnessed a surge of highly heterogeneous computing platforms composed of specialized execution units. This trend is also driven by the heterogeneity of the workloads that execute on those computing platforms, which come hand in hand with performance, efficiency and reliability constraints pertaining to specific application \textit{domains}. So the net picture that has emerged involves domain-specific applications running on domain-specific systems on a chip (SoC).

Domain-specific SoCs expose a variety of options for application task (processes, threads) execution, including but not limited to general-purpose cores, graphics processing units (GPUs), hardware accelerators, application-specific integrated circuits (ASICs), and digital signal processors (DSPs). Due to this degree of heterogeneity, task scheduling becomes less trivial when compared to homogeneous counterparts. In its simplest form, the scheduler can \textit{statically} map application tasks to fixed execution units in the chip (e.g. FFT functions always run on FFT accelerators) to reduce the complexity associated with dynamic (run-time) scheduling decisions. This approach, however, can limit the scheduler's capabilities in making dynamic decisions that can lead to better performance or efficiency. For example, if all the FFT accelerators are already in use, should the scheduler wait until one of them is vacated? Or would it be more convenient to execute the FFT function on a less optimal unit (e.g. a general-purpose core) to reduce its waiting time? Clearly, task scheduling can result in a potentially complex problem in the presence of hardware heterogeneity. Therefore, it becomes critical to provide the right set of tools for early-stage prototyping and evaluation of scheduling algorithms (``policies'') in heterogeneous systems, enabling enough flexibility and exploration space coverage. To the best of our knowledge, such tools are not openly available today or do not provide enough flexibility for early-stage study of the problem.

This paper presents STOMP (\textbf{S}cheduling \textbf{T}echniques \textbf{O}ptimization in heterogeneous \textbf{M}ulti-\textbf{P}rocessors), a queue-based discrete-event simulator that enables fast implementation and evaluation of task scheduling policies in multi-core/multi-processor systems~\cite{stomp}. It implements a convenient interface to allow users and researchers to ``plug in'' new scheduling policies in a simple manner and without the need to interact with STOMP's internal code. We conceive STOMP with the following three goals in mind:

\begin{itemize}
\setlength\itemsep{0pt}
\item \textit{Flexibility}: it is straightforward to define simulated platforms and applications, and to test new scheduling policies through STOMP's ``plug and play'' approach.
\item \textit{Ease of use}: STOMP's default execution mode allows users to quickly configure and run simulations at the right level of abstraction. It also supports more detailed simulation capabilities for expert users as well.
\item \textit{Openness}: the tool is open source and publicly available~\cite{stomp}.
\end{itemize}

STOMP's core queue-based operation approach builds upon the QUTE framework~\cite{5749737}. However, STOMP introduces radically new elements to support the evaluation of heterogeneous SoCs, allowing users to easily configure multi-core/multi-processor systems with varying degrees of heterogeneity. In one of its execution modes, STOMP can be fed with applications represented as directed acyclic graphs (DAGs), which can be either generated in a synthetic manner or from the characterization of real workloads.

STOMP is thoroughly validated against analytical model counterparts (closed-form expressions). For system utilization levels between 10\%-90\%, the average relative errors for steady-state analysis of waiting times are: 0.50\%, 0.83\%, and 1.45\% (for one, two and three servers, respectively).

The rest of the paper is organized as follows: Section~\ref{stomp_simulator} presents STOMP along with its fundamental elements and simulation capabilities. Section~\ref{methodology_and_validation} discusses the strategy and methodology followed to validate the tool. Section~\ref{evaluation} evaluates a set of illustrative scenarios that exercise some of the most relevant features of STOMP. Finally, the related work and conclusions are presented in Sections~\ref{related_work} and~\ref{conclusions}, respectively.
 
\section{STOMP Simulator}
\label{stomp_simulator}

\begin{figure}[!ht]
\centering
  \includegraphics[width=1.00\columnwidth]{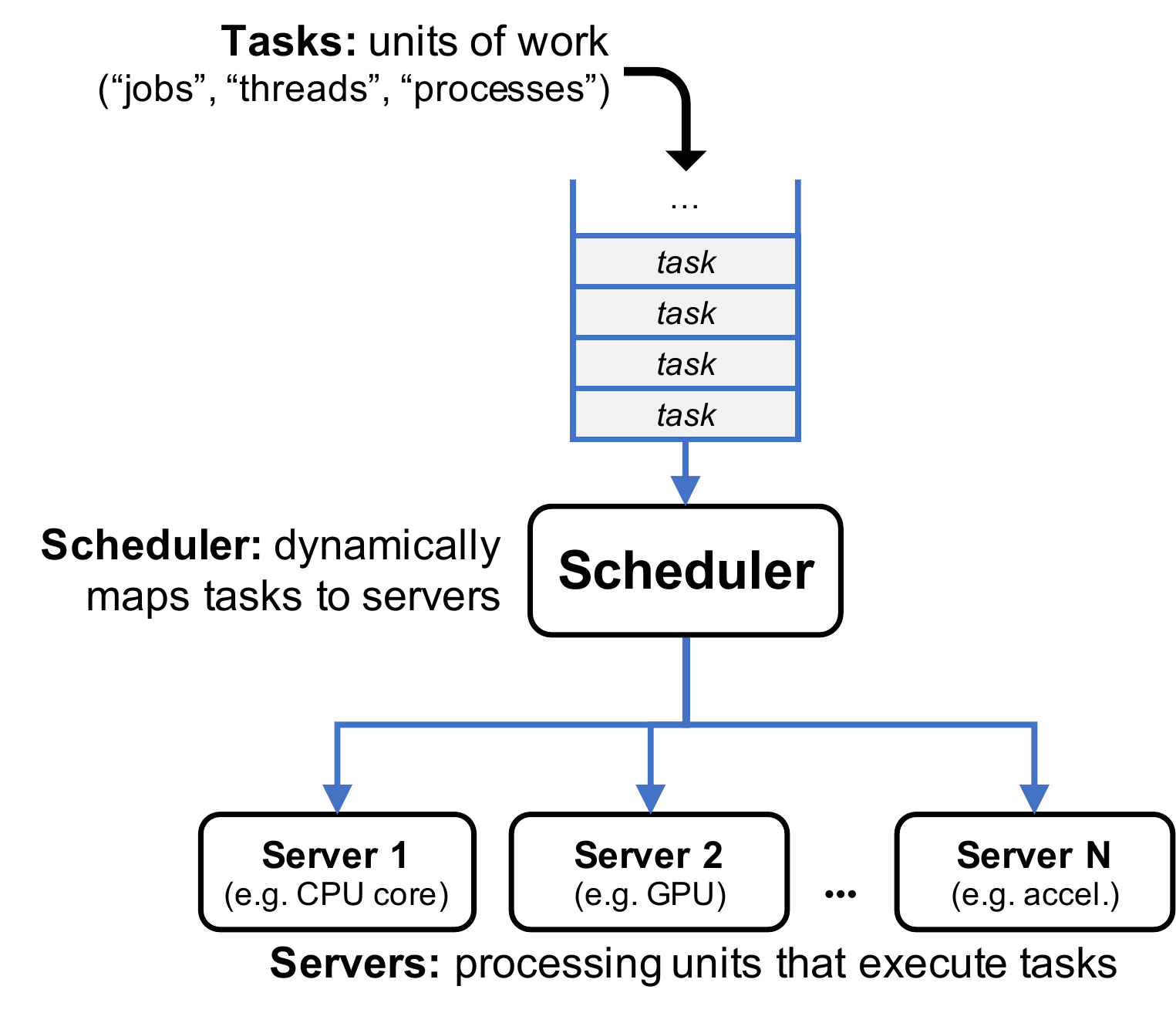}
  \caption{STOMP overview.}
\label{fig.stomp_overview}
\end{figure}

Figure~\ref{fig.stomp_overview} presents a high-level view of the relevant components, including the main \textbf{task queue}, the \textbf{scheduler} and the heterogeneous \textbf{servers} (processing elements). STOMP supports two execution modes: \textit{probabilistic} or \textit{realistic}. In probabilistic mode, the arrival rate and service times (execution times) of tasks are determined by configurable probability distributions (e.g. exponentially for the arrival rate). In realistic mode, the tasks and their associated characteristics (like arrival and service times) are loaded from a trace file provided by the user and previously generated, for example, using real profiling data.

Once inserted in the queue, each task has a set of associated attributes, the following being the most relevant ones:

\begin{itemize}
\setlength\itemsep{0pt}

\item \textit{Target servers}: list of servers (processing elements) where the task can execute on and the order of preference. For example, the list \textit{\{accelerator, GPU, CPU core\}} indicates that the scheduler should first try to place the task in a corresponding accelerator. If an accelerator is not available, then other supported architectures are GPU and CPU core, in that order of preference. It is important to mention that tasks do not necessarily support all the available processing elements –-- e.g. some tasks may only run on CPU cores, or CPU cores \textit{and} GPUs, etc.

\item \textit{Service time}: the list of target servers includes corresponding service (execution) times for each specified processing element. These are \textit{mean} service times used to generate task execution times during simulation. They are ignored in \textit{realistic} execution mode (i.e. when tasks are read from external traces).

\item \textit{Power consumption}: similarly, the list of target servers includes corresponding power consumption information for each specified processing element. This information can be used for implementation of power-aware scheduling policies.

\item \textit{Execution deadline}: single value associated with the task that indicates the amount of time available for execution, and intended for simulation of real-time constrained applications.

\end{itemize}

The user can also specify and configure the characteristics of the servers (processing elements). Obvious options include CPU cores, GPUs, and hardware accelerators. At present, STOMP does not support multi-threaded processing elements --- in other words, once a task is allocated to a processing element, no other task(s) can be scheduled on it until the currently running task finishes its execution.

\subsection{Configuration Parameters}
\label{configuration_parameters}

STOMP simulations are configured through a single JSON file. Some of the most relevant parameters are described below:

\begin{itemize}
\setlength\itemsep{0pt}

\item \texttt{sched\_policy\_module}: indicates the scheduling policy to use. For example, \texttt{policies.test} uses a policy implemented in the \texttt{test.py} file within the \texttt{policies} folder. The implementation of new scheduling policies is discussed in Section~\ref{plug_n_play_scheduling_policies}.

\item \texttt{max\_tasks\_simulated}: maximum number of simulated tasks. Only valid with STOMP's \textit{probabilistic} mode.

\item \texttt{mean\_arrival\_time}: mean task arrival time. Only valid with STOMP's \textit{probabilistic} mode.

\item \texttt{arrival\_time\_scale}: a constant factor to scale \texttt{mean\_arrival\_time}. For example, a 0.5 value will double the task arrival rate (since it halves \texttt{mean\_arrival\_time}); while a 2.0 value will halve the task arrival rate (since it doubles \texttt{mean\_arrival\_time}). Only valid with STOMP's \textit{probabilistic} mode.

\item \texttt{servers}: definition of servers (processing elements) simulated in the system. For example, the following JSON fragment configures a simulated platform with eight general-purpose cores, two GPUs and one FFT accelerator:

\begin{verbatim}
"servers": {
        "cpu_core" : {
            "count" : 8
        },
        "gpu" : {
            "count" : 2
        },
        "fft_accel" : {
            "count" : 1
        }
}
\end{verbatim}

\noindent A server's name is just an arbitrary string and does not assign any specific characteristics to the server. Instead, execution times and power consumption values are part of each task's information.

\item \texttt{tasks}: definition of tasks simulated in the system. Only valid with STOMP's \textit{probabilistic} mode. For example, the following JSON fragment creates a simulated FFT task with specific mean service times and associated standard deviations for the simulated heterogeneous platform:

\begin{verbatim}
"tasks": {
        "fft" : {
            "mean_service_time" : {
                "cpu_core"  : 500,
                "gpu"       : 100,
                "fft_accel" : 10
            },
            "stdev_service_time" : {
                "cpu_core"  : 5.0,
                "gpu"       : 1.0,
                "fft_accel" : 0.1
            }
        }
}
\end{verbatim}

\noindent The standard deviation controls the dispersion of the service (execution) time --- in other words, it allows the user to set the level of determinism of a task's execution characteristics.

\item \texttt{input\_trace\_file}: trace of tasks used for simulation with STOMP's \textit{realistic} mode. The trace also includes the task's arrival time and service times across the different server types in the system. 

\end{itemize}

\textbf{It is important to mention that the concept of ``time'' in STOMP is unitless.} The user is responsible for providing meaning to the time values used in the configuration file --- e.g. a mean service time of ``500'' units of time could mean 500 $\mu$s, or 500 ms, etc.

\subsection{``Plug \& Play'' Scheduling Policies}
\label{plug_n_play_scheduling_policies}

The scheduler module in STOMP (Figure~\ref{fig.stomp_overview}) is responsible for assigning enqueued tasks to servers (processing elements) using a user-specified scheduling policy. The policies can go from very simple decision logic all the way to complex and potentially more ``intelligent'' ones --- eventually using machine learning techniques or other mechanisms for dynamic improvement of the scheduling activities.  In STOMP, new policies are constructed by implementing the abstract class \texttt{BaseSchedulingPolicy}, shown below:

\begin{lstlisting}[frame=single,language=Python,columns=fullflexible]
class BaseSchedulingPolicy:
    
  __metaclass__ = ABCMeta

  @abstractmethod
  def init(self, servers, stomp_stats, stomp_params): pass
    
  @abstractmethod
  def assign_task_to_server(self, sim_time, tasks): pass

  @abstractmethod
  def remove_task_from_server(self, sim_time, server): pass

  @abstractmethod
  def output_final_stats(self, sim_time): pass
\end{lstlisting}

Specifically, the task scheduling decision logic is defined within the \texttt{assign\_task\_to\_server()} method. The user also has the opportunity to implement initialization and finalization activities as part of the \texttt{init()} and \texttt{remove\_task\_from\_server()} methods, and to provide policy-specific statistics via \texttt{output\_final\_stats()} to be displayed at the end of the simulation. One possible (illustrative) example of an \texttt{assign\_task\_to\_server()} implementation is shown below (in this example, the task is only scheduled to the fastest server type, if available):

\begin{lstlisting}[frame=single,language=Python,columns=fullflexible]
def assign_task_to_server(self, sim_time, tasks):

  if (len(tasks) == 0):
    # There aren't tasks to serve
    return None    
        
  # Determine task's best scheduling option
  target_server_type =
      tasks[0].mean_service_time_list[0][0]
                
  # Look for an available server
  for server in self.servers:
    
    if (server.type == target_server_type
       and not server.busy):

      # Assign task in queue's head to server
      server.assign_task(sim_time, tasks.pop(0))
      return server
                
  return None
\end{lstlisting}

Strictly speaking, scheduling policies are implemented as Python modules, with each new policy in a different Python file. The user indicates the module to load (i.e. the policy to use) through the \texttt{sched\_policy\_module} parameter, as we explain in Section~\ref{configuration_parameters}. STOMP's GitHub repository~\cite{stomp} includes examples of scheduling policies that can be used as templates to generate new ones.

\section{STOMP Validation}
\label{methodology_and_validation}

Since STOMP is a queue-based simulator, we address its validation by comparing it against its analytical model counterparts (closed-form expressions). Specifically, we focus on the M/M/k system, as defined by Kendall's notation~\cite{wiki:kendalls_notation}. An M/M/k system models a single queue with $k$ servers (processing elements), where both arrival and service (computation) times are exponentially distributed. In practice, service times in STOMP are normally distributed; however there are only crude approximations for the M/G/k case which are relatively accurate only for a few constrained cases~\cite{gupta_2010,begin:hal-00864323}. This leads us to opt for the M/M/k system to validate the dynamics of the core STOMP simulation engine.

\begin{figure}[!ht]
\centering
  \includegraphics[width=1.00\columnwidth]{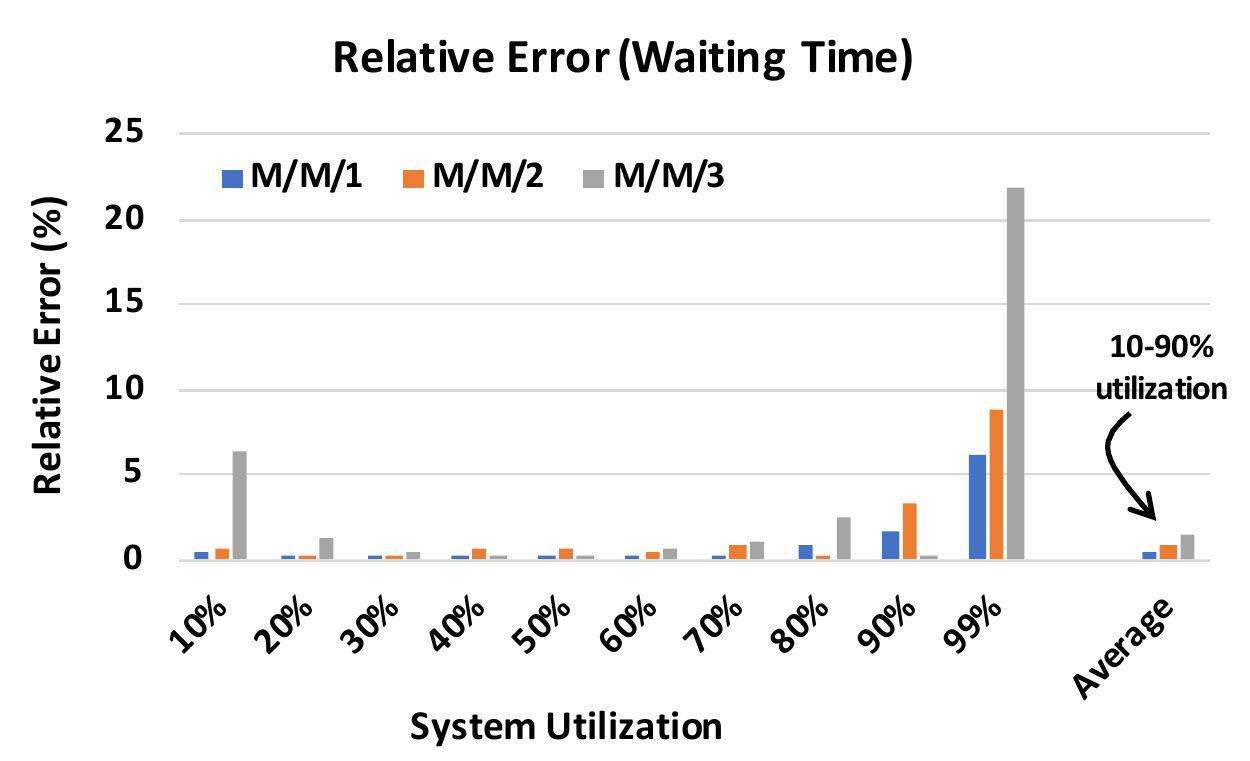}
  \caption{Relative error of the average (steady state) waiting time as a function of the system utilization.}
\label{fig.relative_error_waiting_time}
\end{figure}

Figure~\ref{fig.relative_error_waiting_time} presents the relative error of the average waiting time (steady state analysis) as a function of the system utilization, for the 1-, 2- and 3-server cases (M/M/1, M/M/2 and M/M/3, respectively). The relative error is computed as $\lvert W_{STOMP} - W_{M/M/k} \rvert / W_{M/M/k}$, where $W_{STOMP}$ is the steady-state waiting time generated by STOMP after simulating 1M tasks and $W_{M/M/k}$ is the corresponding waiting time generated using the closed-form formula. In most cases, the relative errors are low. Specifically, for utilization levels between 10\%--90\%, the average relative errors are 0.50\% for M/M/1, 0.83\% for M/M/2, and 1.45\% for M/M/3. The relative error increases for the 99\%-utilization case; it is well known that these formulas are usually not adequate when utilization approaches 100\% as the system becomes less stable~\cite{begin:hal-00864323}.

\begin{figure}[!ht]
\centering
  \includegraphics[width=1.00\columnwidth]{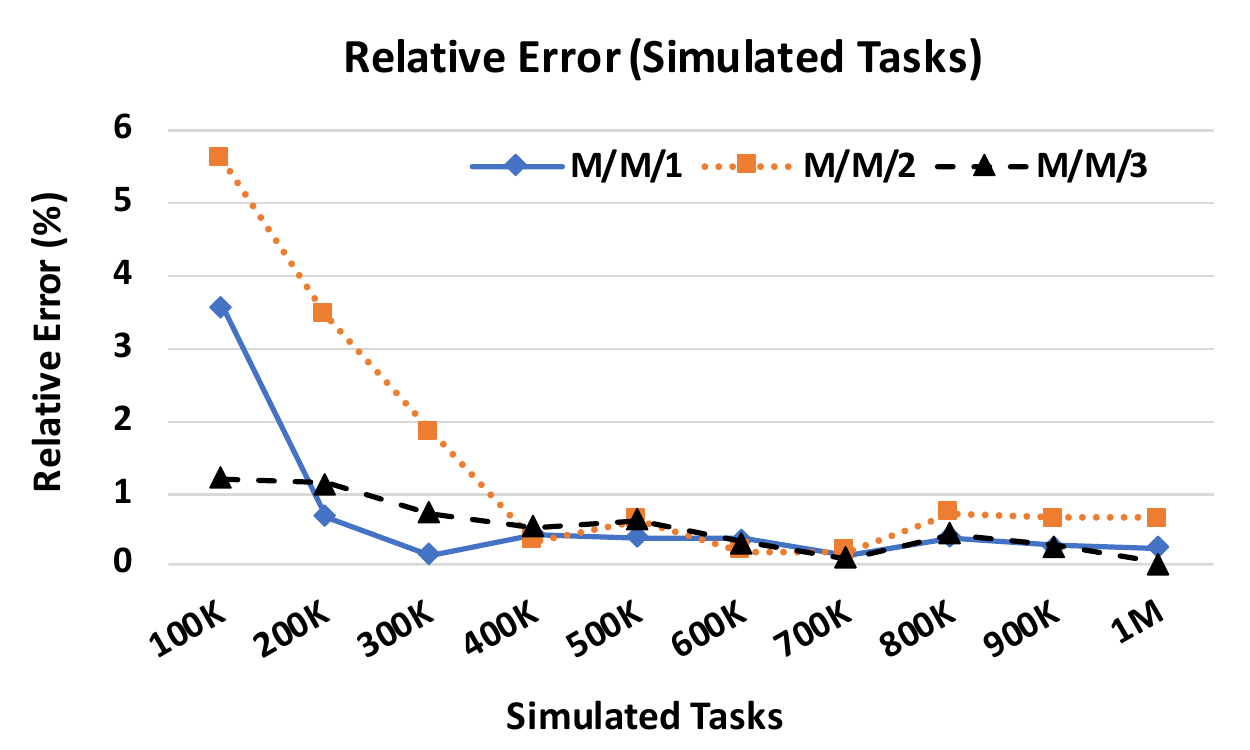}
  \caption{Relative error of the average (steady state) waiting time as a function of the number of simulated tasks (for the 50\% system utilization case).}
\label{fig.relative_error_task_count}
\end{figure}

The accuracy of discrete-event simulators like STOMP is highly dependent on the number of simulated tasks. Usually, when the number of simulated tasks (or ``customers'') is not ``large enough'', the system suffers from instability and warming-up conditions that can invalidate the average (steady state) results. Figure~\ref{fig.relative_error_task_count} presents the relative error of the average waiting time (steady state analysis) as a function of the number of simulated tasks, for the three cases under consideration (M/M/1, M/M/2 and M/M/3). The simulations correspond to 50\% system utilization. As we can observe, the relative error decreases when more tasks are simulated. The ``right'' amount of tasks also depends on the case: 200K tasks is enough to ensure an error smaller than 1\% in the M/M/1 case; while at least 400K and 300K tasks are needed in the M/M/2 and M/M/3 cases, respectively, to ensure the same error bound. In the validation campaign conducted in this work (Figure~\ref{fig.relative_error_waiting_time}), we simulated 1M tasks in all cases to conservatively avoid any possible transient state instabilities during the simulations.

We are currently working on extending this validation analysis to cases where tasks and servers (processing elements) can be of different types (heterogeneous systems). Due to the lack of closed-form formulas for heterogeneous queueing systems, the validation strategy for these cases will require alternative approaches which may imply the use of third-party (already-validated) discrete-event simulators for the heterogeneous cases.
\section{Evaluation}
\label{evaluation}

\begin{figure}[!ht]
\centering
  \includegraphics[width=1.00\columnwidth]{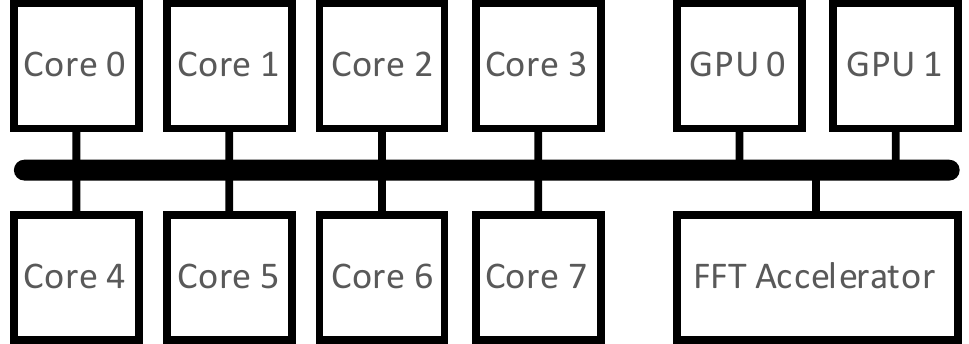}
  \caption{Simulated heterogeneous SoC, including eight general-purpose cores, two GPUs and one FFT accelerator.}
\label{fig.simulated_soc}
\end{figure}

This section evaluates five scheduling policies (available in STOMP's GitHub repository~\cite{stomp} on a simulated SoC platform using STOMP's \textit{probabilistic} mode. This heterogeneous chip, depicted in Figure~\ref{fig.simulated_soc}, consists of eight general-purpose cores, two GPUs and one FFT accelerator. The simulated application is a trace of randomly generated FFT and decoder kernels, which execution times on the simulated SoC are normally distributed with the means specified in Table~\ref{table.execution_times}. We control the dispersion of these execution times by setting the standard deviation to 1\%, 5\% and 50\% of their mean values. These parameters are specified as part of each task's JSON configuration block, as described in Section~\ref{configuration_parameters}. Tasks arrivals are modulated using an exponential distribution, with the following mean values: 50, 75 and 100 units of time.

\begin{table}[ht]
\centering
\caption{Execution time mean values.}
\begin{tabular}{lccc}
\cline{2-4}
\multicolumn{1}{l|}{}                  & \multicolumn{1}{c|}{\textbf{Core}}       & \multicolumn{1}{c|}{\textbf{GPU}}        & \multicolumn{1}{c|}{\textbf{FFT Accel.}}     \\ \hline
\multicolumn{1}{|l|}{\textbf{FFT}}     & \multicolumn{1}{c|}{500}                 & \multicolumn{1}{c|}{100}                 & \multicolumn{1}{c|}{10}                  \\ \hline
\multicolumn{1}{|l|}{\textbf{Decoder}} & \multicolumn{1}{c|}{200}                 & \multicolumn{1}{c|}{150}                 & \multicolumn{1}{c|}{N/A}                 \\ \hline
\end{tabular}
\label{table.execution_times}
\end{table}

The five evaluated scheduling policies are described next:

\begin{itemize}
    \item \textbf{Version 1:} this policy tries to schedule the task in the head of the queue \textit{only} in its best scheduling option (fastest processing element). If the best scheduling option is not available, the task remains in the queue, blocking the rest of the tasks.
    \item \textbf{Version 2:} it tries to schedule the task in the head of the queue in its best scheduling option (fastest processing element). If the best scheduling option is not available, the policy will try to schedule the task in a gradually less-optimal processing element.
    \item \textbf{Version 3:} the policy computes the remaining time for all the processing elements (taking into account currently running tasks), and schedules the head of the queue in the one with the smallest remaining time.
    \item \textbf{Version 4:} similar to version 3, except that if the processing element selected by the policy is currently busy, then the task remains in the queue and the policy will try to schedule the next one in the queue, within a given window of tasks, in a non-blocking fashion.
    \item \textbf{Version 5:} similar to version 4, except that when the scheduler tries to schedule the $i^{th}$ task in the queue, it also factors in the remaining times of all the preceding tasks in the queue, in a non-blocking fashion.
\end{itemize}

Table~\ref{table.configurations} summarizes the most relevant parameters that define the different configurations. A reference JSON configuration file is provided in Appendix~\ref{stomp_configuration_file}, which is updated with the parameters listed in this table to generate the different scenarios.

\begin{table}[ht]
\centering
\caption{Simulated configurations.}
\begin{tabular}{|l|c|}
\hline
\textbf{Parameter}        & \textbf{Values}                                                                                    \\ \hline \hline
\textbf{Simulation Mode}  & Probabilistic                                                                                      \\ \hline
\textbf{Simulated SoC}    & \begin{tabular}[c]{@{}c@{}}8 $\stimes$ general-purpose cores\\
                             2 $\stimes$ GPUs\\
                             1 $\stimes$ FFT accelerator\end{tabular} \\ \hline
\textbf{Simulated Tasks}  & \begin{tabular}[c]{@{}c@{}}100,000\\ (FFT, decoder)\end{tabular}                              \\ \hline
\textbf{Policies}         & Versions 1--5~\cite{stomp}                                                                                              \\ \hline
\textbf{Exec. Time Stdev} & 1\%, 5\%, 50\%                                                                                     \\ \hline
\textbf{Mean Arrival Times} & \begin{tabular}[c]{@{}c@{}}50, 75, 100\\ (units of time)\end{tabular}                              \\ \hline
\end{tabular}
\label{table.configurations}
\end{table}

\begin{figure}[!ht]
\centering
  \includegraphics[width=1.00\columnwidth]{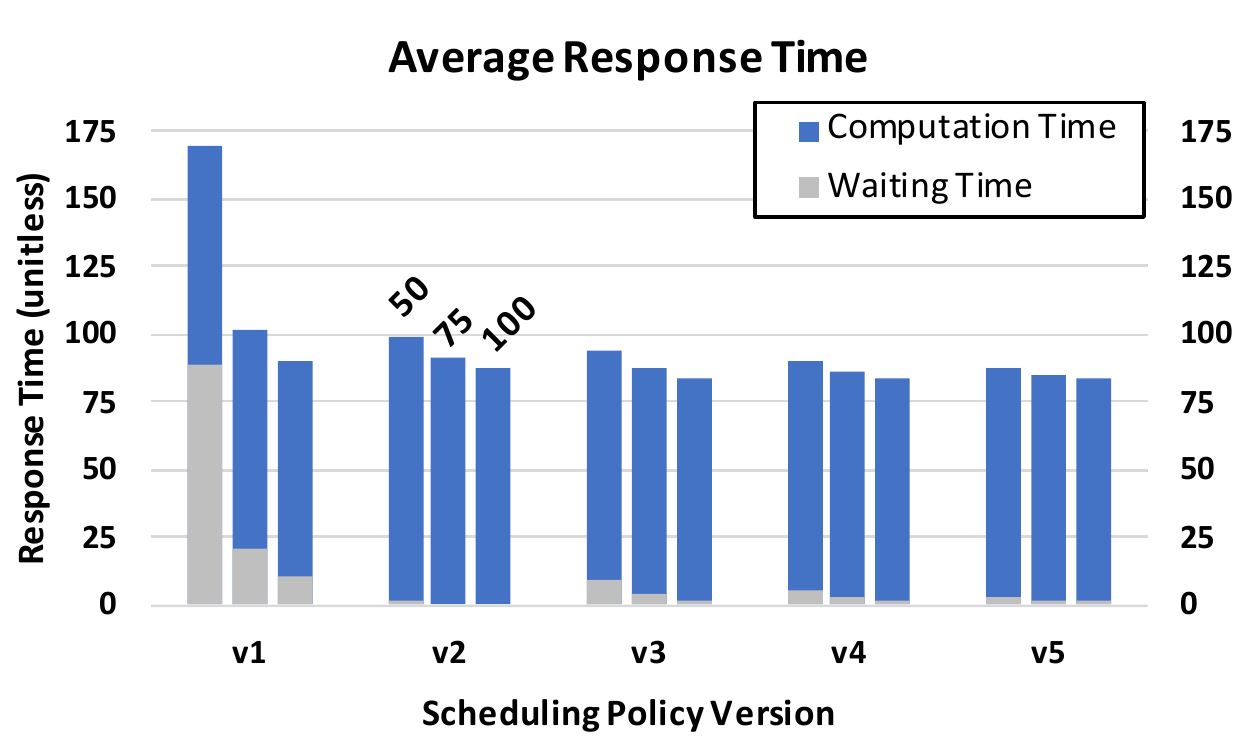}
  \caption{Response time for policies v1--v5 as a function of the arrival time.}
\label{fig.response_time_over_arrival_scale}
\end{figure}

\begin{figure}[!ht]
\centering
  \includegraphics[width=1.00\columnwidth]{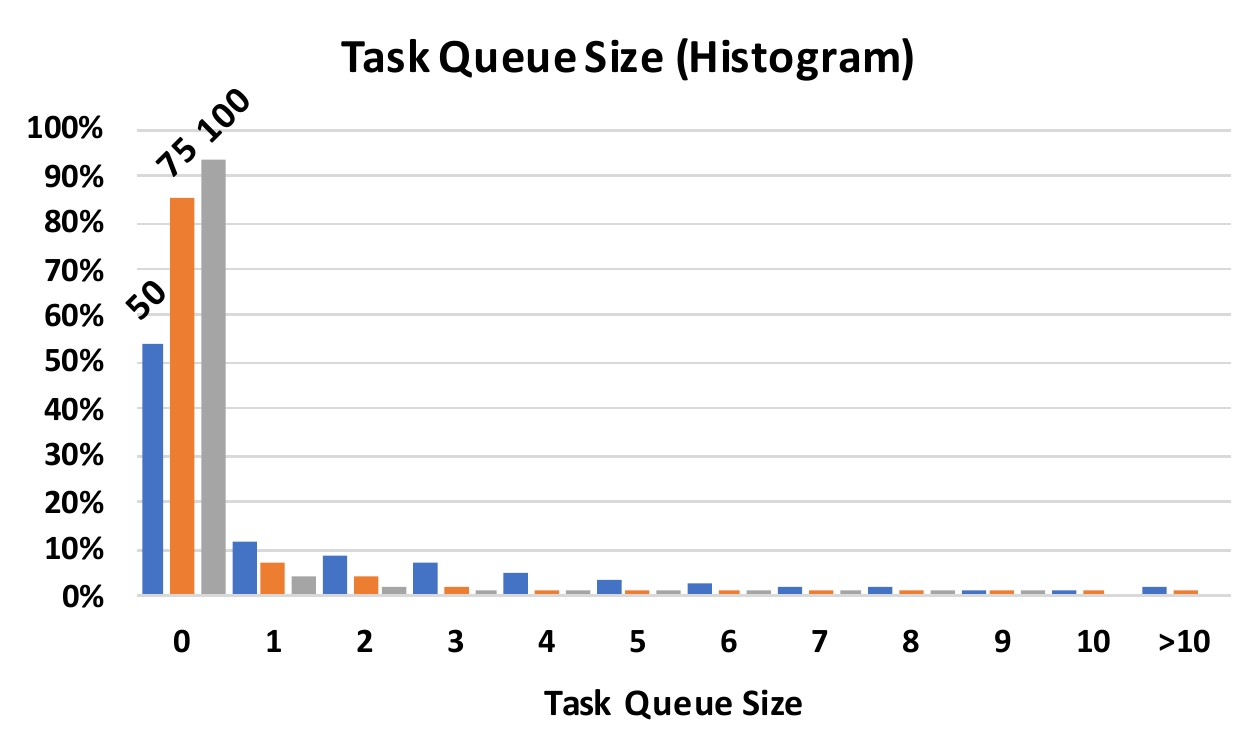}
  \caption{Task queue size histogram as a function of the mean arrival time.}
\label{fig.queue_size_histogram}
\end{figure}

STOMP generates a rich set of output statistics after each simulation. For the sake of illustration, next we focus on \textit{average response time}. It is defined as the time a task spends in the system (since its arrival to its completion) for the entire simulation and averaged across task types. Average response time includes \textit{waiting time} (while the task was in the queue) and \textit{computation time} (while the task was running on its assigned processing element). Figure~\ref{fig.response_time_over_arrival_scale} presents the average response time for the five policy versions (v1--v5) and for different mean arrival times: 50, 75 and 100 units of time. In general, we observe that average response time decreases with larger arrival times, since the system is less ``busy'' and tasks have the opportunity to get a processing element quicker. Across policies, the non-blocking ones (v4 and v5) exhibit slightly better performance due to smaller waiting times. Policy v1, on the other hand, results in large waiting times, especially for smaller arrival times --- in this case, task blocking is more frequent due to the logic of the policy and, therefore, tasks remain longer in the queue. This behavior can be studied using STOMP's generated histograms. For example, Figure~\ref{fig.queue_size_histogram} shows a histogram of the task queue size for policy v1 and for different mean arrival times: 50, 75 and 100 units of time. Clearly, the smaller the arrival time, the larger the number of tasks that have to wait in the queue. For a mean arrival time of 50 units of time, the queue is empty 54\% of the time; while a mean arrival time of 100 units of time results in the queue being empty 94\% of the time.

\begin{figure}[!ht]
\centering
  \includegraphics[width=1.00\columnwidth]{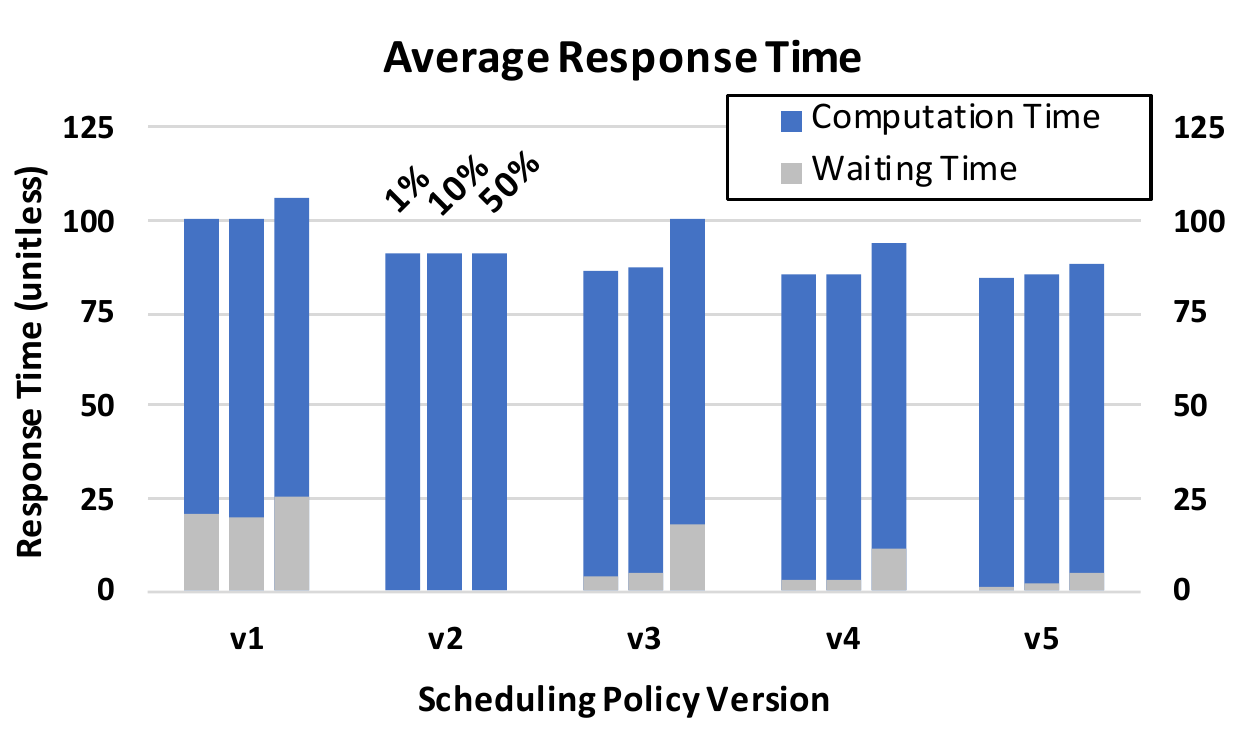}
  \caption{Response time for policies v1--v5 as a function of the computation time dispersion (standard deviation).}
\label{fig.response_time_over_stdev_factor}
\end{figure}

Figure~\ref{fig.response_time_over_stdev_factor} is equivalent to Figure~\ref{fig.response_time_over_arrival_scale}, but varying the dispersion (standard deviation) of the computation time. Since the computation (service) times are normally distributed, changing the dispersion does not affect the average computation time given that we run our simulations ``long enough.'' Some policies ``suffer'' more from higher computation time dispersion. This is the case of policies v3 and v4, which rely heavily on estimating remaining times of running tasks. When the dispersion is large (e.g. 50\%), these estimations are not accurate enough, leading the policies to make less efficient scheduling decisions. Policy v5, although similar to v3 and v4, softens this issue by adopting a more effective way to estimate remaining times.

In addition to response time, computation time, waiting time, and queue size histograms, STOMP generates a variety of output statistics including the utilization of the processing elements in the simulated SoC, and timing analysis per processing element type and per task type.
\section{Related Work}
\label{related_work}

A plethora of work exists on various scheduling policies for heterogeneous architectures. However, most of these scheduling policies are either evaluated on in-house simulators or runtime systems. To the best of our knowledge, STOMP is the first open-source tool for agile evaluation of scheduling algorithms in heterogeneous system that was conceived with ``plug \& play'' flexibility in mind.

Runtime systems like StarPU~\cite{augonnet2009starpu} and Nanos++~\cite{nanos-1,nanos-2} provide scheduler frameworks. However it can be extremely tedious to develop and compare scheduling policies in a runtime system that has been developed to perform various kernel operations. Moreover the data structures and models of the runtime system can constrict the generality of a scheduling policy being developed \cite{augonnet2009starpu}.

TaskSim~\cite{rico2010scalable} is a simulator for the execution of tasks on decoupled accelerator systems with the capability of scheduling tasks. DS3 is a simulator for heterogeneous SoCs with scheduling and power management features~\cite{2003.09016}. Both TaskSim and DS3 are full-system simulators and, consequently, agile evaluation and comparison of multiple scheduling algorithms is not necessarily a straightforward process.
\section{Conclusion}
\label{conclusions}

The auspicious power-performance efficiency benefits of heterogeneous chip multiprocessors come with new challenges, being task (process, thread) scheduling one of the most critical ones. In this context, it becomes decisive to count with the proper tools for early-stage evaluation of scheduling policies. In this paper, we present STOMP (\textbf{S}cheduling \textbf{T}echniques \textbf{O}ptimization in heterogeneous \textbf{M}ulti-\textbf{P}rocessors), a simulator for fast implementation and evaluation of task scheduling policies in domain-specific systems on a chip (SoC)

STOMP provides a convenient interface for ``plugging'' in new scheduling policies in a simple manner. Similarly, the user can define the characteristics of the modeled SoC and the simulated application, either in \textit{probabilistic} or \textit{realistic} (trace-based) modes. At the end of a simulation run, STOMP generates a rich set of output statistics which can be conveniently used for deeper understanding of the simulation dynamics. STOMP exhibits small relative errors when validated against closed-formed equivalent models during steady-state analysis.

To the best of our knowledge, STOMP covers a domain which has received little attention --- that of agile simulation tools for early-stage evaluation of scheduling policies in domain-specific SoCs.

\section*{Acknowledgments}

This research was developed with funding from the Defense Advanced Research Projects Agency (DARPA). The views, opinions and/or other findings expressed are those of the authors and should not be interpreted as representing the official views or policies of the Department of Defense or the U.S. Government. This document is approved for public release: distribution unlimited.
\bibliographystyle{IEEEtranS}
\bibliography{references}

\newpage
\onecolumn

\appendices
\section{STOMP Configuration File}
\label{stomp_configuration_file}

This is a reference JSON configuration file for STOMP, also available in STOMP's GitHub repository (stomp.json)~\cite{stomp}.

\small
\begin{verbatim}
    {
      "general" : {
          "logging_level":       "INFO",
          "random_seed":         0,
          "working_dir":         ".",
          "basename":            "",
          "pre_gen_arrivals":    false,
          "input_trace_file":    "",
          "output_trace_file":   ""
      },
  
      "simulation" : {
          "sched_policy_module": "policies.simple_policy_ver3",
          "max_tasks_simulated": 100000,
          "mean_arrival_time":   50,
          "power_mgmt_enabled":  false,
          "max_queue_size":      1000000,
          "arrival_time_scale":  1.0,

          "servers": {
              "cpu_core" : {
                  "count" : 8
              },
              "gpu" : {
                  "count" : 2
              },
              "fft_accel" : {
                  "count" : 1
              }
          },

          "tasks": {
              "fft" : {
                  "mean_service_time" : {
                      "cpu_core"  : 500,
                      "gpu"       : 100,
                      "fft_accel" : 10
                  },
                  "stdev_service_time" : {
                      "cpu_core"  : 5.0,
                      "gpu"       : 1.0,
                      "fft_accel" : 0.1
                  }
              },

              "decoder" : {
                  "mean_service_time" : {
                      "cpu_core"  : 200,
                      "gpu"       : 150
                  },
                  "stdev_service_time" : {
                      "cpu_core"  : 2.0,
                      "gpu"       : 1.5
                  }
              }
          }
      }
    }
\end{verbatim}

\end{document}